\documentclass[a4paper,10pt]{article}

\usepackage[width=18cm,height=26cm]{geometry}
\usepackage{amsmath,amssymb}
\usepackage{graphicx}
\usepackage{multicol}
\usepackage[square,numbers,sort&compress]{natbib}
\usepackage{doi}
\usepackage{hyperref}
\hypersetup{colorlinks,pdftitle={A New Method to Optimize Finite Dimensions Thermodynamic Models:
application to an Irreversible Stirling Engine},pdfauthor={F.~Lanzetta,
A.~Vaudrey,
P.~Baucour},citecolor=blue,filecolor=blue,linkcolor=blue,urlcolor=blue}

\title{\bf A New Method to Optimize Finite Dimensions Thermodynamic Models:
application to an Irreversible Stirling Engine}
\author{F.~Lanzetta\thanks{University of Franche-Comte, FEMTO-ST, UMR CNRS 6174,
Parc technologique, 2 avenue Jean Moulin, 90000 Belfort, France.} \and
A.~Vaudrey\thanks{Corresponding author :
\href{mailto:alexandre.vaudrey@ecam.fr}{alexandre.vaudrey@ecam.fr},
\href{http://orcid.org/0000-0002-8613-774X}{\tt ORCID iD : 0000-0002-8613-774X},
ECAM Lyon, University of Lyon, LabECAM/P\^ole \'Energ\'etique, 40 mont\'ee Saint-Barth\'elemy,
69321, Lyon, France.} \and P.~Baucour\textsuperscript{*}}

\begin{document}
\maketitle

\begin{abstract}
	Different economical configurations, due for instance to the relative
	cost of the fuel it consumes, can push a heat engine into operating whether at
	\emph{maximum efficiency} or at \emph{maximum power} produced. Any
	relevant design of such system hence needs to be based, at least partly,
	on the knowledge of its specific "power vs.~efficiency" characteristic
	curve. However, even when a simple model is used to describe the engine,
	obtained for example thanks to Finite Dimensions Thermodynamics, such
	characteristic curve is often difficult to obtain and takes an explicit
	form only for the simplest of these models. When more realistic models
	are considered, including complex internal subsystems or processes, an
	explicit expression for this curve is practically impossible to obtain.
	In this paper, we propose to use the called Graham's scan algorithm in
	order to directly obtain the power vs.~efficiency curve of a realistic
	Stirling engine model, which includes \emph{heat leakage},
	\emph{regenerator effectiveness}, as well as \emph{internal} and
	\emph{external irreversibilities}. Coupled with an adapted optimization
	routine, this approach allows to design and optimize the same way simple
	or complex heat engine models. Such method can then be useful during the
	practical design task of any thermal power converter, almost regardless
	to its own internal complexity.

	\medskip


\end{abstract}

\begin{multicols}{2}

\section{Introduction}

\emph{Air Engines}, i.e. \emph{Stirling} and \emph{Ericsson} ones, can help to solve
various energy problems. They can be supplied by different kinds of heat sources
(combustion, solar or nuclear energy) and they are usually quiet, at least when
compared to Internal Combustion Engines.  Benefiting from these good points,
such engines can thus be included for instance in domestic \emph{cogeneration},
i.e. Combined Heating and Power systems \cite{ATE-033-0119,Energy-049-0229} or
in \emph{trigeneration}, i.e. Combined Cooling, Heating and Power
systems~\cite{AE-102-1303}. Stirling engines can also be used for example in
solar-power plants~\cite{RSER-007-0131}, in solar-pumping
systems~\cite{RE-081-0319} or in combined cycle with other machines
\cite{RSER-062-0089}, as e.g. high temperature fuel cells
\cite{Energy-061-0087,VPPC-REM-2010}. 

Air engines operate over \emph{closed} and \emph{regenerative} thermodynamic
cycles, with typically a permanent gas, as e.g. helium or air as working fluid.
It means that, among other complex components, very specific heat exchangers
have to be designed in order to exchange the required heat quantities between
the working gas and the hot and cold sources of the engine; but also from this
gas to itself at a different moment of the cycle, thanks to a
\emph{regenerator}~\cite{TBJHS-007-0259}. 

Before the use of a complete but complex engine model --- such as those
presented for example in \cite{Dyson-2004,Energy-076-0445} --- which requires a
lot of information sometimes not already available, a simpler model, based for
example on \emph{Finite Dimensions Thermodynamics}
\cite{JNET-024-0327,JAPh-079-1191-1218} can help to design such systems. This
kind of model can thus be used for sort of preliminary designs, for example to
obtain a first draft of the relation between the power produced by the engine
and its corresponding efficiency, so the usually called "power vs. efficiency"
or $(\eta,\dot{W})$ characteristic curve.

Such curve is really important, because once it is known, the final selected
trade-off between a maximum efficiency and a maximum power produced by the whole
engine must be obtained as the result of an economical calculation. Broadly
explained, the higher the cost of the fuel consumed, the higher the required
efficiency for the whole system, although the higher its size and so its
investment cost. Indeed, for a heat engine for example, increasing its
efficiency means decreasing its internal irreversibilities by decreasing, among
other things, the temperature differences within its heat exchangers, and so the
sizes of the latter. At the contrary, the lower the cost of the fuel, the higher
the power produced in order to decrease the investment cost as far as possible,
as already explained by Novikov~\cite{Novikov1958} in 1958 for
nuclear plants. The knowledge, for a given engine, of its $(\eta,\dot{W})$
curve, is thus needed to complete successfully such optimization
step~\cite{ECM-036-0001}. 

Unfortunately, an explicit algebraic expression of such curve, of the form
$\dot{W} = f(\eta)$ for example, can be obtained and drawn only for the simplest
models of engines. If the complexity of the latter increases, for example
because including further subsystems or processes, as regenerator or heat leaks
for example, the aforementioned curve can only be obtained in an implicit way,
i.e. through the use of the complete model. $\dot{W}$ and $\eta$ are then two
different functions of several parameters related to the engine's design and to
its operating conditions. Anyhow, without any dedicated method, this curve is
nothing bu easy to obtain. 

In this paper, we present a new approach, based on the called Graham's
algorithm, that allows to directly obtained the required characteristic curve
with a minimal number of calls for the engine's model. This approach will be
applied to a model of Stirling engine, including different complex subsystems
and phenomena.

\section{Finite dimension thermodynamics of Stirling engines}

Although the search for engine models dealing with maximum rates of useful
energy is not really new \cite{JNET-021-0239,JNET-039-0199}, such idea have been
actually widely spread only after the publication of the seminal paper from
\citeauthor{Curzon1975}~\cite{Curzon1975} in 1975. Since then, the method commonly
called \emph{Finite Time Thermodynamics} or \emph{Finite Dimensions
Thermodynamics}, have been applied to the optimization of numerous
different energy systems \cite{JNET-024-0327,JAPh-079-1191-1218}, including
obviously the Stirling engine. 

\citeauthor{RGT-032-0509}~\cite{RGT-032-0509} studied the basic Stirling cycle
composed by definition of four strokes : one isochoric compression, one
isochoric expansion and two isothermal heat exchanges. Irreversibilities were
supposed to occur during the heat exchange processes only, and were due to the
transfer of a heat rate through a difference between two constant temperatures,
the ones of the hot and cold thermal reservoirs and the one of the working gas.
\citeauthor{Blank1994}~\cite{Blank1994} added a regenerator to a similar model
of Stirling engine, the former being considered as a usual countercurrent heat
exchanger of limited effectiveness. In another paper, the same authors also
introduced possible radiative heat exchange processes with the working gas in
considering the same model propelled by solar energy \cite{IJAE-015-0131}.
\citeauthor{Popescu1996}~\cite{Popescu1996} generalized such model in
incorporating heat leaks phenomena and an non adiabatic regenerator of limited
effectiveness. \citeauthor{Energy-019-0837}~\cite{Energy-019-0837} took into
account the motion of the working gas within the Stirling engine and then
related the performances of the whole engine to its rotational speed.
\citeauthor{JFI-330-0967}~\cite{JFI-330-0967}, after obtaining a general
relation for the maximum \emph{mechanical efficiency} of such heat engine cycle,
considering the effect of friction, applied this relation to the specific case
of the Stirling engine. He considered limited values for the heat rates
exchanged by the working gas and a heat leak between the hot and cold part of
the engine~\cite{IJR-022-0991}. \citeauthor{JAPh-084-2385}~\cite{JAPh-084-2385}
generalized the same approach to a wide family of Stirling-like engine cycles.
Over the years, more sophisticated Stirling engine models have been developed
using Finite Dimension Thermodynamics, including the effects of pressure drop
due to fluid friction inside the engine~\cite{ECM-040-1723}, inefficiencies of
compression and expansion strokes~\cite{IJR-023-0863}, presence of dead
volumes~\cite{RE-031-0345} or even actual geometries of
engines~\cite{OGST-066-0747}. It is worth noting that in all these studies, the
power vs.~efficiency curve was expressed as an explicit relation $\dot{W} =
f(\eta)$ of a variable complexity.

\section{Theoretical model}

\subsection{Heat transfer law}

The form of the used heat transfer law influences the results obtained in the
modeling of any thermodynamic system. The simplest expression for the heat transfer
law is the linear one, that can be used to describe \emph{conductive} or
\emph{convective} heat transfer processes: \begin{equation}
	\dot{Q} = h \cdot A \cdot \Delta T = K \cdot \Delta T
	\label{eq:Newton_law}
\end{equation} Where $h$ is the heat transfer coefficient, $A$ a characteristic
area, $K = h \cdot A$ a thermal conductance and $\Delta T$ a temperature
difference.

The linear law \eqref{eq:Newton_law} is in good agreement with reality for
laminar and turbulent convection and have been widely applied to study
endoreversible and irreversible models of machines. However, for systems
operating at high temperatures, \emph{radiation} is the major heat transfer
process. The latter is described by the Stefan-Boltzmann law, expressed for two
bodies at respective temperatures $T_{1}$ and $T_{2}$, as: \begin{equation}
	\dot{Q} = F \cdot A \cdot \sigma \cdot \left(T_{1}^{4}-T_{2}^{4}\right)
	\label{eq:SB_Law}
\end{equation} With $F$ the form factor including emittance of the two bodies
and view factors, and the Stefan-Boltzmann constant $\sigma=5.67 \cdot
10^{-8}\,{\rm W \cdot m}^{-2} \cdot {\rm K}^{-4}$ \cite{Sigel-Howell}. Many
thermodynamic models, including radiation, have been utilized for internal and
external combustion engines, gas turbine cycles and power plants. When a system
houses convective as well as radiative heat exchange processes, the Dulong-Petit
law may also be useful~: \begin{equation}
	\dot{Q} = k \cdot A \cdot \Delta T^{n} \label{eq:Dulong_law} 
\end{equation} With $k$ and $n$ two specific coefficients which depend on the
nature of the heat transfer $n=5/4$ or $n=4/3$ for a natural laminar, or
turbulent convection, on a plate respectively \cite{AJPh-058-0956}. The
Dulong-Petit law \eqref{eq:Dulong_law} have been applied to describe heat
transfer in models of heat engines and refrigerators \cite{JAPh-074-2216}.

\subsection{Thermodynamic cycle}

\begin{figure*}
	\centering
	\includegraphics[width=0.6\textwidth]{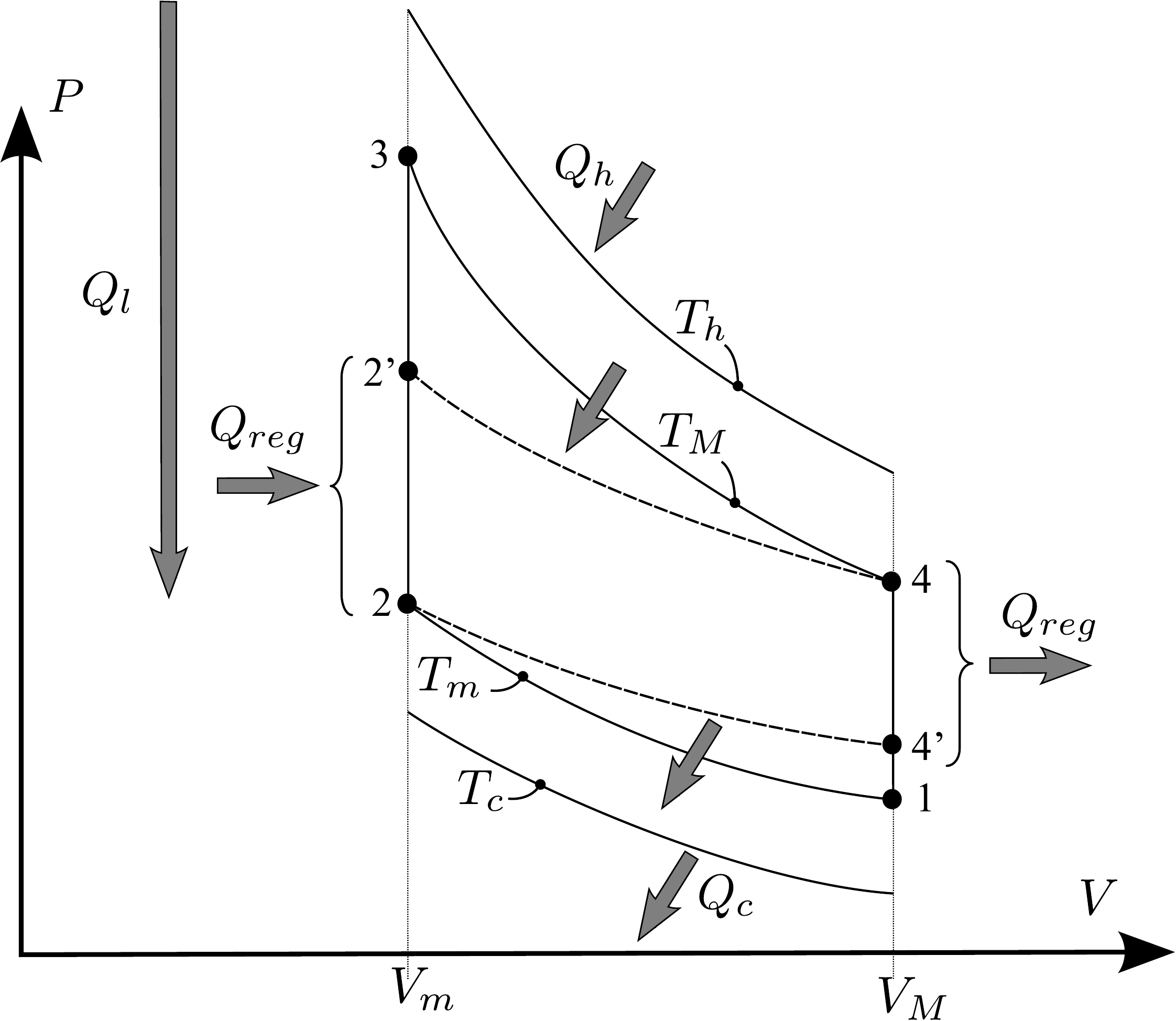}
	\caption{Representations of the Stirling cycle under study in the
	$(V,p)$ diagram.}
	\label{fig:CyclePV}
\end{figure*}

\begin{figure*}
	\centering
	\includegraphics[width=0.6\textwidth]{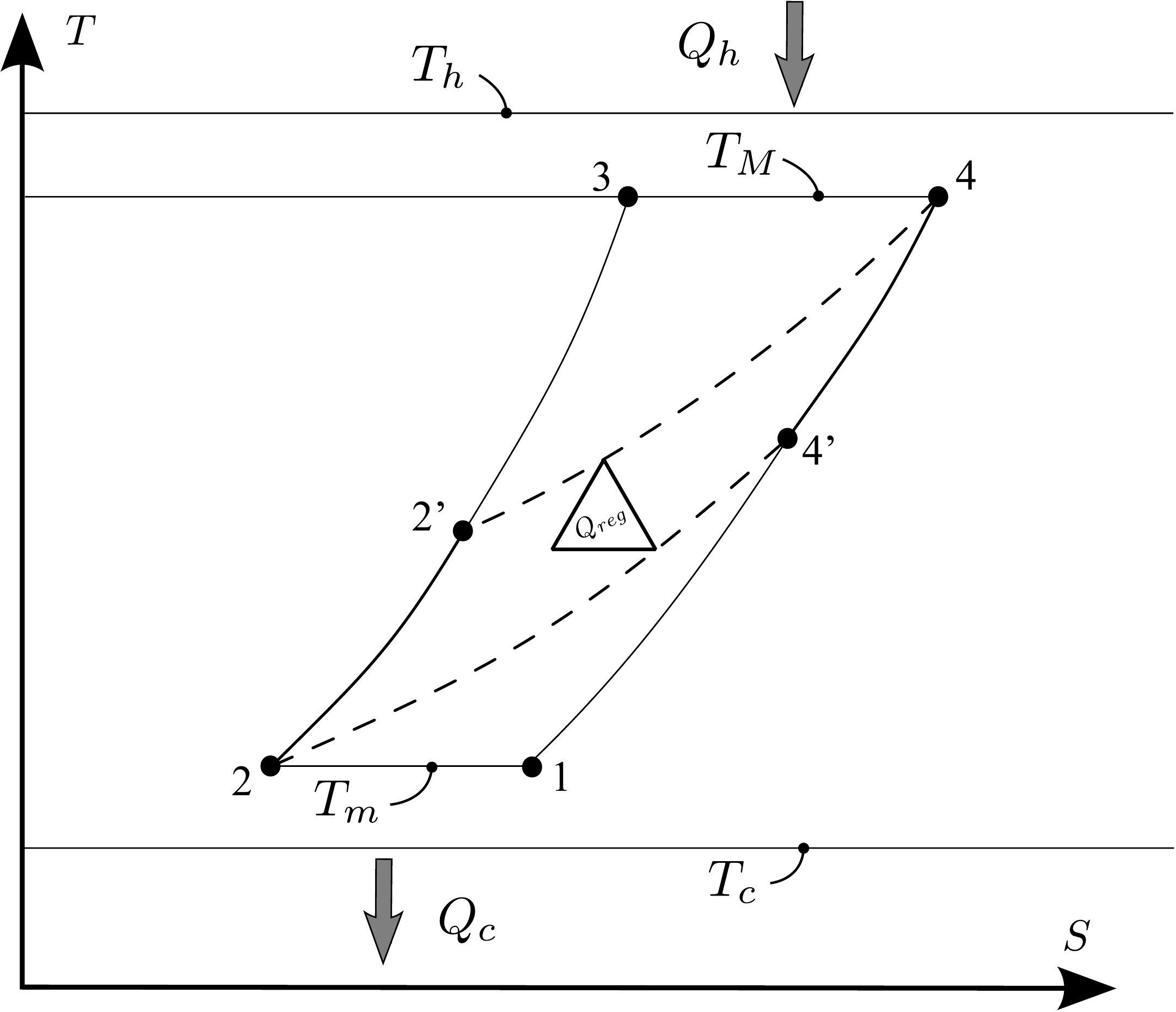}
	\caption{Representations of the Stirling cycle under study in the
	$(S,T)$ one. $T_h$ and $T_c$ are respectively the temperatures of the
	hot and cold heat sources of the engine, while $T_M$ and $T_m$ the
	maximum and minimum ones reached by the working gas, respectively.
	Thanks to a regenerator, a part $Q_{44'}$ if the heat $Q_{41}$ rejected
	by the gas during its cooling process ($4-1$) is salvaged by the same
	fluid in order to heat up itself from $T_2$ to $T_{2'}$. The remaining
	heat quantity $Q_{2'3}$ required to reach the high temperature $T_3 =
	T_M$ is supplied by the heat source at temperature $T_h$.}
	\label{fig:CycleTS}
\end{figure*}

The $(V,p)$ and $(S,T)$ diagrams of a Stirling-like externally and internally
irreversible heat engine are shown in Figures~\ref{fig:CyclePV} and
\ref{fig:CycleTS}, respectively.  The cycle operates between a heat source of
constant temperature $T_h$ and a cold sink of constant temperature $T_c$. This
cycle consists of two isothermal processes ($1-2$ and $3-4$) and
two isochoric processes ($2-3$ and $4-1$). During the heat rejection process
($1-2$), heat flows from the working gas to a heat sink across the finite
temperature difference $T_m - T_c$ and during the heat addition process ($3-4$),
heat is transferred from the high temperature source across the finite
difference temperature $T_h-T_M$. These two processes are thus irreversible.

In the present study, the Stirling engine is subjected to significant losses due
to heat leaks, directly from the hot source to the cold sink. \begin{figure*}
\begin{centering}
	\includegraphics[width=0.7\textwidth]{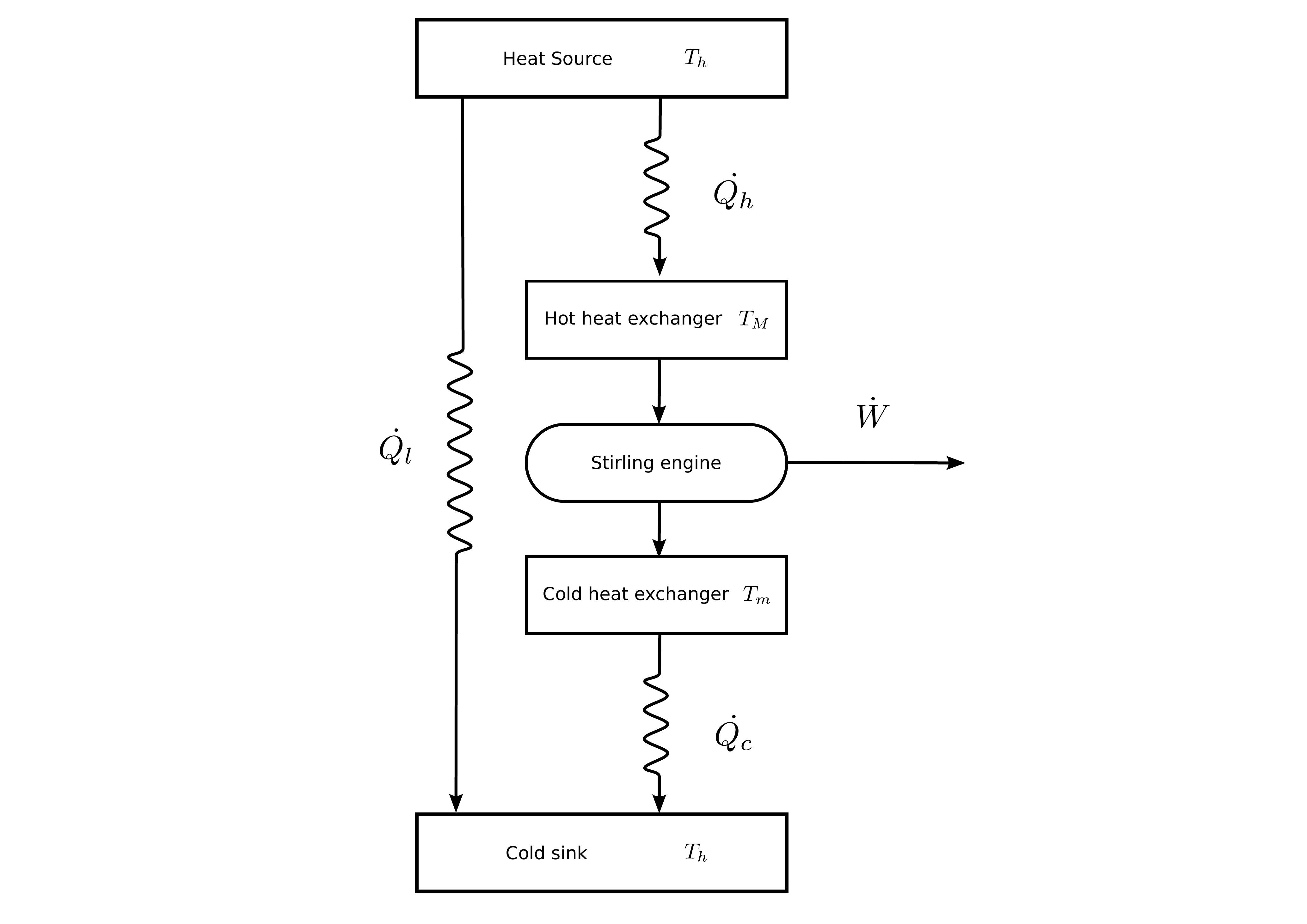}
	\caption{Schematic diagram of the system studied. The hot heat source at
	temperature $T_h$ supplies the heat rate $\dot{Q}_h$ to the Stirling
	engine through a hot heat exchanger which raise the working gas
	temperature at a maximum value $T_M$. The engine also rejects a lower
	heat rate $\dot{Q}_c$ to the cold sink at temperature $T_c$ through a
	cold heat exchanger which makes the gas temperature decrease to $T_m$.
	$\dot{Q}_l$ is the heat rate which leaks directly from the hot to the
	cold heat reservoir.} 
	\label{fig:System}
\end{centering}
\end{figure*} The schematic diagram in Figure~\ref{fig:System} presents the only
sizable irreversibilities taken to be heat leak between hot source and cold sink
and finite rate transfer from them to the engine.  All of these heat transfer
processes utilize a linear model, similar as \eqref{eq:Newton_law} because of
its simplicity and because this model is as general as possible regarding to the
usual practical temperature ranges of such engines.

\subsection{Internal irreversibilities}

Let us consider now the internal irreversibilities occurring inside the engine
itself. They are due to : \begin{itemize}
	\item Imperfect heat exchange inside the regenerator with the working
		gas.
	\item Heat leak by conduction heat transfer between the hot and cold
		reservoirs.
	\item Heat resistances between the working gas and the heat
		reservoirs.
	\item Flow frictions between the fluid and the inner walls of the
		engine.
\end{itemize}

\subsubsection{Regenerator with effectiveness} \label{paragraph-regenerator}

\paragraph{The regenerator} A regenerator operates as a thermal storage element,
alternately absorbing and releasing heat with the working gas
\cite{TBJHS-007-0259,Reader1983,Walker1993,Organ1992}. In an ideal, i.e.
perfectly regenerated cycle, all of the heat rejected by the working gas in
order to heat up the regenerator during the constant volume phase ($4-1$) allows
to later heat up the same fluid from its low to its high temperature during the
cooling process of the regenerator ($2-3$), see Figure~\ref{fig:CycleTS}. The
regenerator is usually designed of a porous media in order to provide the
largest contact surface with the working gas. The long-awaited characteristics
of such regenerative matrix are: its high heat capacity; its large contact area;
its small porous matrix which minimizes the flow losses and, in minimizing dead
spaces, improves the ratio of maximum to minimum pressure. In the rest, the heat
quantities exchanged with the working gas are supposed to be so by a realistic non
adiabatic regenerator, of effectiveness $\alpha$~\cite{Tlili2012}.

\paragraph{The regenerator effectiveness} In general, the effectiveness defines
how well a real heat exchanger is performing regarding to an ideal one operating
across the same temperature difference~\cite{Shah-Sekulic}. In our case, the
\emph{regenerator effectiveness} $\alpha$ is thus defined as the ratio of the
heat $Q_{44'}$ rejected (or absorbed $Q_{22'}$) by the working gas to the
maximum heat $Q_{41}$ rejected (or absorbed $Q_{23}$) during the isothermal
process ($4-1$) (or ($2-3$)), as presented in Figures~\ref{fig:CyclePV} and
\ref{fig:CycleTS}. 

Effectiveness $\alpha$ will be henceforth considered to be less than one but
with a constant value during the heating and cooling periods, contrary to
proposals of different authors \cite{Blank1994,Blank1996,Blank1999}. In reality,
the gas flow rate is indeed not constant and then the heat transfer coefficient
between the solid matrix and the gas varies during each process of the cycle. In
such situation, the efficiency cannot be constant during the cycle. Such
imperfect regeneration have been modeled for example by
\citeauthor{Feng1998}~\cite{Feng1998} and the regenerator effectiveness defined
by the ratio of effective and maximal energy stored or released during
regenerative process~\cite{Lanzetta1996,Nika1997,Nika1995}. 

\paragraph{Mathematical model of the regenerator} For the heating period of the
regenerator, the effectiveness is thus: \begin{equation}
	\alpha = \frac{Q_{44'}}{Q_{41}} = \frac{m \cdot c_v \cdot
	(T_{4'}-T_M)}{m \cdot c_v \cdot (T_m-T_M)} \leq 1 
	\label{eq:Effectiveness}
\end{equation} Then, the working gas at the outlet of the heating period of the
regenerator is: \begin{equation}
	T_{4'}=T_M-\alpha \cdot (T_M-T_m) \label{eq:T4prime}
\end{equation} The heat quantity rejected by the gas during process ($4-1$) of
duration $t_{41}$ can be expressed in two different ways: \begin{align} 
	Q_{41} & = m \cdot c_{v} \cdot (T_m-T_M) \label{eq:t410}
	\\
& = K_{\text{reg}} \cdot (T_m-T_M) \cdot t_{41}
	\label{eq:t41} 
\end{align} This amount of heat can be split in two parts
$Q_{41}=Q_{44'}+Q_{4'1}$ with, according to \eqref{eq:Effectiveness}, the one
not stored by the regenerator: \begin{equation}
	Q_{4'1}=(1-\alpha) \cdot Q_{41} \label{eq:Q4prime}
\end{equation} 

For the cooling period of the regenerator, the same effectiveness is expressed
as: \begin{equation}
	\alpha=\frac{Q_{22'}}{Q_{23}} = \frac{m \cdot c_{v} \cdot
	(T_{2'}-T_m)}{m \cdot c_{v} \cdot (T_M-T_m)} \label{eq:Effectiveness-2}
\end{equation} And the regenerator cooling period outlet temperature is:
\begin{equation}
	T_{2'}=T_m+\alpha \cdot (T_M-T_m) \label{eq:T2prime}
\end{equation} The heat absorbed by the gas during the process ($2-3$) of
duration $t_{23}$ is then: \begin{align}
	Q_{23} & = m \cdot c_v \cdot (T_M-T_m) \label{eq:Q23_TM_Tm0}\\ 
	& = m \cdot \frac{r}{\gamma-1} \cdot (T_M-T_m)
	\label{eq:Q23_TM_Tm} \\ 
	& = K_{\text{reg}} \cdot (T_M-T_m) \cdot t_{23} \label{eq:t23} 
\end{align} This amount of heat can also be expressed as
$Q_{23}=Q_{22'}+Q_{2'3}$ with: \begin{equation}
	Q_{2'3} = (1-\alpha) \cdot Q_{23} \label{q23}
\end{equation} the part not supplied by the regenerator which, using equation
\eqref{eq:Q23_TM_Tm}, can also be written as: \begin{equation}
	Q_{2'3} = (1-\alpha) \cdot m \cdot \frac{r}{\gamma-1} \cdot (T_M-T_m)
	\label{eq:Q2p3}
\end{equation} We can also notice that, considering equations \eqref{eq:t410},
\eqref{eq:Q4prime}, \eqref{eq:Q23_TM_Tm0} and \eqref{q23}, we have the equality:
\begin{equation}
	Q_{2'3} = \vert Q_{4'1} \vert \label{heat-equality}
\end{equation}

\subsubsection{Heat leak}

The heat leak exists because of the stationary heat transfer at a rate
characterized by a thermal conductance $K_l$, occurring during the overall cycle
duration $t_{\text{cycle}}=t_{12}+t_{23}+t_{34}+t_{41}$
\cite{Popescu1996,Bejan1988,Chen1994,Kodal2000}. This process is assumed to be
linear in temperature difference and so: \begin{equation}
	Q_l = K_l \cdot (T_h-T_c) \cdot t_{\text{cycle}} 
\end{equation} As previously explained, the thermal conductance $K_l$
corresponds to the heat transfer area times the overall heat transfer
coefficient based on that area, see equation \eqref{eq:Newton_law}.

\subsubsection{Irreversibility degree of the cycle}

The second law of thermodynamics, applied to an irreversible cycle requires
that: \begin{equation}
	\oint_{\text{cycle}}\frac{\delta Q}{T}<0 
\end{equation} Then, combining both heat transfer processes and their
respective (equivalent) source temperatures, we can write: \begin{equation}
	\left(\frac{Q_M}{T_M}+\frac{Q_{2'3}}{T_{2'3}}\right) - \left(\frac{\vert
	Q_m\vert}{T_m}+\frac{\vert Q_{4'1}\vert}{T_{4'1}}\right) \leq 0
	\label{eq:2nd_Law}
\end{equation} With :
\begin{itemize} 
	\item $Q_M$ the amount of heat received by the working gas during the
		isothermal expansion ($3-4$) occurring at temperature $T_M$:
		\begin{equation}
			Q_M=m \cdot r \cdot T_M \cdot \ln(\epsilon)
			\label{eq:QM}
		\end{equation} Where $\epsilon$ is the \emph{compression ratio}
		$\epsilon=\frac{V_{4}}{V_{3}}=\frac{V_{1}}{V_{2}}$ of the
		engine.
	\item $\vert Q_m\vert$ the absolute value of the amount of heat rejected by
		the gas during the isothermal compression ($1-2$) at temperature
		$T_m$: \begin{equation}
			\vert Q_m\vert= m \cdot r \cdot T_m \cdot
			\ln(\epsilon)
		\end{equation}
\end{itemize} Transformations ($2'-3$) and ($4'-1$) being nothing but
isothermal, see Figure~\ref{fig:CycleTS}, temperatures $T_{2'3}$ and $T_{4'1}$
are mostly of a symbolic nature and must be expressed using the other
parameters of the model. The assumption of a perfect gas makes it possible to
determine their expressions using equations \eqref{eq:T4prime}, \eqref{eq:t410}
and \eqref{eq:Q4prime}, and introducing the temperature ratio $\theta=T_M/T_m$.
Starting with $T_{4'1}$, we obtain: \begin{align} 
	T_{4'1} & = \frac{Q_{4^{\prime}1}}{\Delta S_{4'1}} =
	\frac{(1-\alpha) \cdot m \cdot c_v \cdot (T_m-T_M)}{m \cdot c_v \cdot \ln(T_m/T_4')}
	\nonumber \\
	& = T_m \cdot \frac{(1-\alpha) \cdot (\theta-1)}{\ln\left( \theta-\alpha
	\cdot (\theta-1)\right)} \label{eq:T4p1}
\end{align} The same approach can be applied to the other equivalent temperature
$T_{2'3}$, using in this case equations \eqref{eq:T2prime} and \eqref{eq:Q23_TM_Tm0} :
\begin{align}
	T_{2'3} & = \frac{Q_{2'3}}{\Delta S_{2'3}} = \frac{(1-\alpha) \cdot m
	\cdot c_v \cdot (T_M-T_m)}{m \cdot c_v \cdot \ln(T_M/T_{2'})} \nonumber \\
	& = T_m \cdot
	\frac{(1-\alpha) \cdot (\theta-1)}{\ln\left( \frac{\theta}{1+\alpha
	\cdot (\theta-1)}\right)} \label{eq:T2p3}
\end{align} An internal irreversibility parameter $r_{\Delta s}$ can be
introduced, which characterizes the degree of internal irreversibility of the
cycle \cite{Bejan1988,Kodal2000,Lingen1997,Ozkaynak1994,Wu1992,Yan1995} :
\begin{equation} 
	r_{\Delta s}=\frac{S_{4'}-S_{2}}{S_{4}-S_{2'}} \geq 1 \label{eq:R_ds}
\end{equation} Indeed, the ratio $r_{\Delta s}$ represents the ratio of two entropy
differences of the working gas during its heat exchange with the hot source
only, so : \begin{equation}
	S_4 - S_{2'} = \frac{Q_M}{T_M} + \frac{Q_{2'3}}{T_{2'3}}
	\label{eq:Sdiffh}
\end{equation} for the heating process of the gas ($2'-4$), and :
\begin{equation}
	S_{4'}-S_2 = \frac{\vert Q_m \vert}{T_m} + \frac{\vert Q_{4'1}
	\vert}{T_{4'1}} \label{eq:Sdiffc}
\end{equation} for the cooling process of the gas ($4'-2$), see 
Figure~\ref{fig:CycleTS}. $r_{\Delta s}$ will take into account all the
irreversibilities except those due to heat losses. For $r_{\Delta s}=1$, the
heat engine is endoreversible whereas for $r_{\Delta s}>1$, it is internally
irreversible. By substituting equations \eqref{eq:R_ds}, \eqref{eq:Sdiffh},
\eqref{eq:Sdiffc}, we obtain, instead of equation \eqref{eq:2nd_Law} the
following one: \begin{equation}
	\left(\frac{\vert Q_m\vert}{T_m} +
	\frac{\vert Q_{4'1}\vert}{T_{4'1}}\right) - r_{\Delta s} \cdot
	\left(\frac{Q_M}{T_M} + \frac{Q_{2'3}}{T_{2'3}}\right)=0
	\label{eq:2nd_Law-2}
\end{equation} Equation \eqref{eq:2nd_Law-2} has to be solved in order to determine
for example $\vert Q_m\vert$ as a function of the others parameters:
\begin{equation} 
	\vert Q_m\vert = T_m \cdot \left[r_{\Delta s} \cdot
	\left(\frac{Q_M}{T_M} + \frac{Q_{2'3}}{T_{2'3}}\right) -
	\frac{\vert Q_{4'1}\vert}{T_{4'1}}\right]
	\label{eq:Qm}
\end{equation} Considering that $Q_{2'3}=\vert Q_{4'1}\vert$, according to
\eqref{heat-equality}; using anew the ratio $\theta=\frac{T_M}{T_m}$; and
combining equations \eqref{eq:QM}, \eqref{eq:Q2p3}, \eqref{eq:T4p1} and
\eqref{eq:T2p3}, we obtain: \begin{align}
	\vert Q_m\vert = & T_m \cdot \left(r_{\Delta s} \cdot \frac{Q_M}{T_M}
	+ Q_{2'3} \cdot \left(\frac{r_{\Delta s}}{T_{2'3}} - \frac{1}{T_{4'1}}
	\right) \right) \nonumber \\
	= & m \cdot r \cdot T_m \cdot \Bigg[r_{\Delta s} \cdot \ln\epsilon \nonumber \\
	- & \frac{\ln\left(\left[\theta-\alpha \cdot \left(\theta-1\right)\right] \cdot
	\left[\frac{1+\alpha \cdot \left(\theta-1\right)}{\theta}\right]^{r_{\Delta
	s}}\right)}{\gamma-1}\Bigg] \label{eq:Qm_fTm} 
\end{align}

\subsection{External irreversibilities}

These irreversibilities are due to heat transfer processes between the hot and
cold heat reservoirs and the working gas. All heat transfer phenomena are
assumed to be linear with temperature differences, according to
\eqref{eq:Newton_law}. The heat rate occurring from the hot source at constant
temperature $T_h$ to the working gas at constant temperature $T_M$, with the
thermal conductance $K_h$ during the period $t_{34}$ requested to accomplish
the isothermal expansion ($3-4$) gives: \begin{align} 
	Q_h & = K_h \cdot (T_h-T_M) \cdot t_{34}
	\label{eq:t34}\\ 
	\text{and } Q_h & = Q_M+Q_{2'3} \label{eq:Qh} 
\end{align} The other heat rate from the gas at constant temperature $T_m$ to
the cold sink at constant temperature $T_c$ with the thermal conductance $K_{l}$
during the period $t_{12}$ requested to accomplish the isothermal
compression ($1-2$) gives: \begin{align} 
	\vert Q_c\vert & =  K_c \cdot (T_m-T_c) \cdot t_{12} \label{eq:t12}\\ 
	\text{and } \vert Q_c\vert & = \vert Q_m\vert+\vert Q_{4'1}\vert \label{eq:Qc}
\end{align}

\subsection{Stirling engine power and efficiency}

The total amount of heat transferred from the hot source, noted $Q_{th}$, is,
according to \eqref{eq:Qh}, composed by the sum: \begin{equation} 
	Q_{th}=Q_h+Q_{l}=Q_M+Q_{2'3}+Q_{l} \label{eq:Q_th}
\end{equation} The total amount of heat rejected to the cold sink noted $Q_{tc}$
is, using \eqref{eq:Qc}: \begin{equation}
	Q_{tc}=\vert Q_c\vert + Q_{l}=\vert Q_m\vert+\vert Q_{4'1}\vert+Q_{l}
	\label{eq:Q_tc}
\end{equation} Then, the \emph{net work} $W$ produced by the Stirling engine is,
according to the first law of thermodynamics and to the equations
\eqref{eq:Q_th}, \eqref{eq:Q_tc} and \eqref{heat-equality}: \begin{equation}
	W=Q_{th}-Q_{tc}=Q_M-\vert Q_m\vert \label{eq:W_1st_ver} 
\end{equation} The combination of equations \eqref{eq:W_1st_ver}, \eqref{eq:QM}
and \eqref{eq:Qm_fTm} provides a new expression for work $W$ : \begin{align} 
	W = & m \cdot r \cdot T_m \cdot \Bigg[ \left(\theta-r_{\Delta s}\right)
	\cdot \ln\epsilon \nonumber \\
	+ & \frac{\ln\left( \left[\theta-\alpha \cdot \left(\theta-1\right)\right]
	\cdot \left[\frac{1+\alpha \cdot \left(\theta-1\right)}{\theta}\right]^{r_{\Delta
	s}}\right)}{\gamma-1}\Bigg] \label{eq:Net_Work}
\end{align} The complete cycle period $t_{\text{cycle}}$ is the sum of all the ones of
individual processes, and considering equations \eqref{eq:t12} for ($1-2$),
\eqref{eq:t23} for ($2-3$), \eqref{eq:t34} for ($3-4$) and \eqref{eq:t41} for
($4-1$), we can express the former as: \begin{align} 
	t_{\text{cycle}} = & t_{12}+t_{23}+t_{34}+t_{41} \nonumber \\ 
	= & \frac{\vert Q_c\vert}{K_c \cdot
	(T_m-T_c)}+\frac{Q_{23}}{K_{\text{reg}}
	\cdot (T_M-T_m)} \nonumber \\
	& +\frac{Q_h}{K_h \cdot (T_h-T_M)}+\frac{Q_{41}}{K_{\text{reg}} \cdot
	\left(T_m-T_M\right)}
\end{align} adding equations \eqref{eq:Qc} and \eqref{eq:Qh} and considering
that $Q_{23}=-Q_{41}$, we obtain a new expression for the complete cycle period:
\begin{align} 
	t_{\text{cycle}} & = \frac{\vert Q_m\vert+\vert Q_{4'1}\vert}{K_c \cdot
	(T_m-T_c)}+\frac{Q_M+Q_{2'3}}{K_h \cdot (T_h-T_M)} \nonumber \\
	& + \frac{2 \cdot Q_{23}}{K_{\text{reg}} \cdot (T_M-T_m)} 
\end{align} Combining the previous result with equation \eqref{eq:Qm_fTm}, we
have: \begin{align}
	t_{\text{cycle}} = & m \cdot r \cdot T_m \cdot \Bigg[ \frac{\ln\epsilon +
	\frac{(1-\alpha)\cdot (\theta-1)}{\gamma-1}}{K_c \cdot (T_m-T_c)}
	\label{eq:tcycle} \\
		& + \frac{\theta \cdot \ln\epsilon + \frac{(1-\alpha) \cdot
	(\theta-1)}{\gamma-1}}{K_h \cdot (T_h-T_M)} + \frac{2}{T_m \cdot
	K_{\text{reg}}
	\cdot (\gamma-1)}\Bigg] \nonumber
\end{align} The total duration of the cycle $t_{\text{cycle}}$ characterizes the
capacity of the system to transfer heat from the heat reservoirs towards fluid
in a finite time.

Thus, combining equations \eqref{eq:Net_Work} and \eqref{eq:tcycle}, the net
power produced by the Stirling engine is given by the fraction $\dot{W} =
W/t_{\text{cycle}}$. For a simplification and better visualization of the results, a
new parameter $\Delta$ is introduced, such as: \begin{equation} 
	\begin{cases} 
		\theta= & \frac{T_M}{T_m}\\ 
		\Delta= & T_M-T_m \end{cases}\textrm{ and }\begin{cases} T_M= & \frac{\theta\cdot\Delta}{\theta-1}\\ 
			T_m= & \frac{\Delta}{\theta-1}
		\end{cases} \label{def-delta-theta}
\end{equation} the net power $\dot{W}$ thus becomes an explicit function of
two variables $\left(\theta,\Delta\right)$ : the equation \eqref{eq:Net_Power}
presented on page \pageref{eq:Net_Power}. \begin{figure*}
	\begin{align}
		\dot{W} & =\frac{W}{t_{\text{cycle}}} = \frac{(\theta-r_{\Delta s}) \cdot
		(\gamma-1) \cdot \ln(\epsilon) + \ln\left(\left(\theta-\alpha
		\cdot (\theta-1) \right) \cdot \left( \frac{1+\alpha \cdot
		(\theta-1)}{\theta}\right)^{r_{\Delta s}}\right)}{(\theta-1)
		\cdot \left( \frac{\ln(\epsilon) \cdot (\gamma-1) + (1-\alpha)
		\cdot (\theta-1)}{K_c \cdot \left(\Delta-T_c \cdot
		(\theta-1)\right)} + \frac{\theta \cdot \ln(\epsilon) \cdot
		(\gamma-1) + (1-\alpha) \cdot (\theta-1)}{K_h \cdot \left(T_h
		\cdot (\theta-1)-\theta \cdot
		\Delta\right)}+\frac{2}{K_{\text{reg}}
		\cdot \Delta}\right)} \label{eq:Net_Power} \\
		\eta = & \frac{W}{Q_{th}} = \frac{Q_M-\vert
		Q_m\vert}{Q_M+Q_{2'3}+Q_l} \nonumber \\
		= & \frac{(\theta-r_{\Delta s}) \cdot (\gamma-1) \cdot
		\ln(\epsilon) + \ln\left(\left(\theta-\alpha \cdot
		(\theta-1)\right) \cdot \left(\frac{1+\alpha \cdot
		(\theta-1)}{\theta}\right)^{r_{\Delta s}}\right)}{\theta \cdot
		\ln(\epsilon) \cdot (\gamma-1) + (\theta-1) \cdot
		\left[(1-\alpha)+K_l \cdot (T_h-T_c) \cdot \left[
		\frac{\ln(\epsilon) \cdot (\gamma-1)+(1-\alpha) \cdot
		(\theta-1)}{K_c \cdot \left(\Delta-T_c \cdot (\theta-1)\right)}
		+ \frac{\theta \cdot \ln(\epsilon) \cdot (\gamma-1) + (1-\alpha)
		\cdot (\theta-1)}{K_h \cdot \left( T_h \cdot (\theta-1) - \theta
		\cdot \Delta \right)}+\frac{2}{K_{\text{reg}} \cdot \Delta}\right]
		\right]} \label{eq:Thermal_efficiency}
	\end{align}
\end{figure*} Finally, the thermal efficiency of the Stirling engine cycle $\eta$
also becomes an explicit function of $\left(\theta,\Delta\right)$: the
equation \eqref{eq:Thermal_efficiency} page \pageref{eq:Thermal_efficiency}.

Equations \eqref{eq:Net_Power} and \eqref{eq:Thermal_efficiency} are in
accordance with the expressions previously determined by
\citeauthor{Kaushik2000}~\cite{Kaushik2000,Kaushik2001},
\citeauthor{Tlili2012}~\cite{Tlili2012} and
\citeauthor{Yaqi2011}~\cite{Yaqi2011}. Equations \eqref{eq:Net_Power} and
\eqref{eq:Thermal_efficiency} extend the general framework of Finite Dimensions
Thermodynamic analysis of a Stirling engine with additional parameters such as
$r_{\Delta s}$ , $\alpha$ and $K_{\text{reg}}$ which are not always used in previous
works.

\subsection{Special cases}

\subsubsection{Comparison with the reversible Stirling engine}

From equations \eqref{eq:Net_Power} and \eqref{eq:Thermal_efficiency}, the
perfect regenerative Stirling engine is recovered for: \begin{itemize}
	\item No internal irreversibilities \emph{i.e.} $r_{\Delta s}=1$. The
		internal flow of fluid does not create entropy, for example
		because of fluid friction effects. 
	\item No external irreversibilities \emph{i.e.} $K_{l}=0$ and $K_h$, $K_c$,
		and $K_{\text{reg}}$ tending towards infinity. Perfect heat exchange takes
		place between the exchanger surfaces and the gas, which implies
		instantaneous heat exchanges and a cycle time equal to zero,
		according to equation \eqref{eq:tcycle}.
		This approach assumes that the boundary walls have the same
		temperature as the fluid (\emph{i.e.} $T_M=T_h$ and $T_m=T_c$). 
	\item Perfect regeneration \emph{i.e.} $\alpha=1$. The porous matrix of
		the regenerator exchanges the heat during the different phases
		of the thermodynamic cycle with a perfect effectiveness.
\end{itemize} For all of these conditions, the net work \eqref{eq:Net_Work} and
the efficiency \eqref{eq:Thermal_efficiency} can be simplified as:
\begin{equation} 
	W_{\text{rev}} = m \cdot r \cdot (T_h-T_c) \cdot \ln(\epsilon) \label{eq:Wrev}
\end{equation} and: \begin{equation} 
	\eta_{\text{rev}}=1-\frac{T_c}{T_h}\label{eq:nurev}
\end{equation} The resulting efficiency and net work correspond to the
reversible Stirling engine \cite{Reader1983,Walker1993,Organ1992}. One should
notice that the efficiency of \eqref{eq:nurev} is equal to the one obtained
with a reversible Carnot heat engine \cite{Blank1994,Blank1996,Chen1994}.

\subsubsection{Comparison with the endoreversible Stirling
engine\label{sub:Comp-with-endoreversible}}

The endoreversible Stirling engine affected by the irreversibility of finite
rate heat transfer and perfect regeneration $(\alpha=1)$ corresponds to an
absence of heat leak $(K_{l}=0)$ and an absence of internal irreversibilities
($r_{\Delta s}=1$) \cite{Blank1994,Blank1999,Chen1994}. For all of these
conditions, the net work of equation \eqref{eq:Net_Work} can be simplified as:
\begin{align} 
	W_{\text{endo}} = & \left(\frac{T_m}{K_c \cdot (T_m-T_c) \cdot
	(T_M-T_m)} + \right.\nonumber \\ 
	&+\frac{T_M}{K_h \cdot \left(T_h-T_m\right) \cdot
	\left(T_M-T_m\right)} \nonumber \\
	& \left. +\frac{2}{K_{\text{\text{reg}}} \cdot \ln(\epsilon) \cdot
	(\gamma-1)}\right)^{-1}\label{eq:Wendorev}
\end{align} With the same hypotheses, the efficiency of equation
\eqref{eq:Thermal_efficiency} becomes the one of an endoreversible Stirling
engine at maximum power, so the subscript ``mp'': \begin{equation}
	\eta_{\text{endo,mp}}=1-\sqrt{\frac{T_c}{T_h}}\label{eq:nuChamb}
\end{equation} Expression of \eqref{eq:nuChamb} is independent on the thermal
conductances. It depends only on the temperature of the heat reservoirs
\emph{i.e.} the hot source at $T_h$ and the cold sink at $T_c$. This result
corresponds to the famous thermal efficiency at maximum power, sometimes called
``\textsc{Chambadal-Novikov-Curzon-Ahlborn} efficiency''
\cite{Chambadal1957,Novikov1958,Curzon1975,Chen2001} or more recently
``\textsc{Reitlinger} efficiency'' \cite{JNET-039-0199}. The resulting maximum
power of equation \eqref{eq:Net_Power} is in
this case: \begin{equation}
	\dot{W}_{\text{endo},\max}= \frac{K_c \cdot K_h}{K_c+K_h}
	\cdot \left(\sqrt{T_h}-\sqrt{T_c}\right)^{2}\label{eq:PowerChamb}
\end{equation} Let us now study the complete and irreversible Stirling engine
model.

\section{Numerical analysis of the engine model}

\subsection{Numerical tools}

\begin{figure*}[t]
	\centering
	\includegraphics[width=0.75\textwidth]{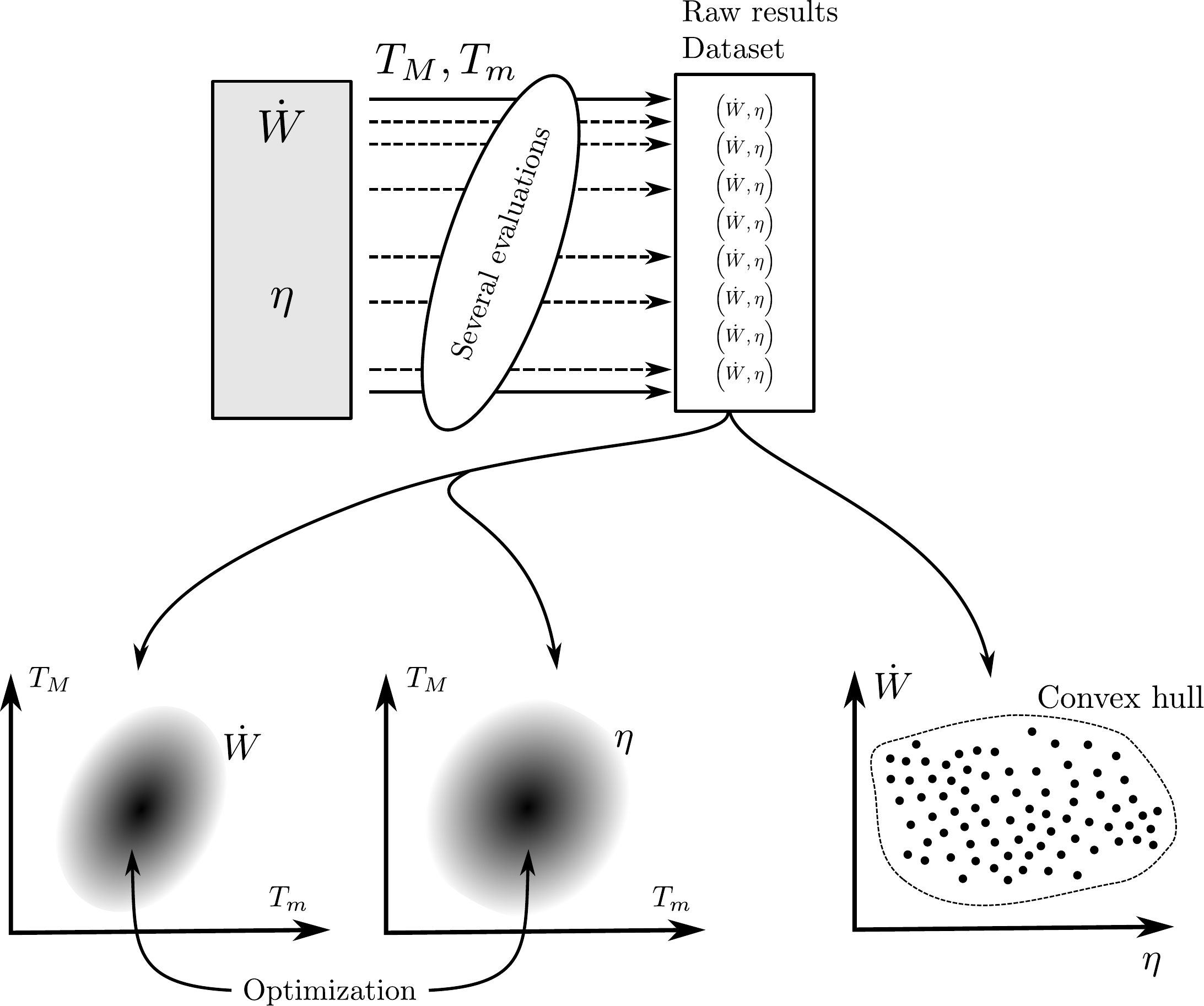}
	\caption{Working framework implemented in Python. Top center: for a
	fixed set of parameters $r_{\Delta s}$, $\gamma$, $\epsilon$, $T_{c}$,
	$T_{h}$, $\alpha$, $K_{h}$, $K_{l}$, $K_{c}$ and $K_{\text{reg}}$,
	several evaluations of $\dot{W}$ and $\eta$, using equations
	\eqref{eq:Net_Power} and \eqref{eq:Thermal_efficiency} respectively, are
	done with different values of $T_M$ and $T_m$, or conversely of $\Delta$
	and $\theta$. Bottom left: once these values are obtained, maximum power
	and efficiency can be computed, each corresponding to a specific values
	of $(T_M,T_m)$ or $(\Delta,\theta)$, as presented in
	Figure~\ref{fig:Power-and-efficiency-surface}. Bottom right: finally,
	both couples of values $(\eta,\dot{W})$ are used to draw the ``power vs.
	efficiency'' curve, thanks to the convex hull algorithm, as presented in
	Figure~\ref{fig:Power-vs-efficiency}.}
	\label{fig:Working-framework}
\end{figure*}

Dealing with large analytical relationships as \eqref{eq:Net_Power} and
\eqref{eq:Thermal_efficiency} impose to use numerical calculations and
optimization algorithms. In the past, several similar studies were published on
optimization of thermodynamic systems, using for instance Genetic Algorithm (GA)
\cite{Ahmadi2015a,Toghyani2014,Ahmadi2013,Ahmadi2013d} or Particle Swarm
Optimization (PSO) \cite{Duan2014,Bert2014}. The main advantages of these
techniques are their ability to deal with a large number of parameters and to
treat multi-objective problems. Furthermore they do not require the
calculation of the objective function's derivative. Their main drawback is
however the readability of the obtained results because: \begin{itemize}
	\item even if such algorithm provide an optimum, it do not provide the
		system behaviour in the vicinity of the optimum.
	\item it do not show where the optimum is localized along the power
		vs.~efficiency curve.
\end{itemize} In this paper a framework is proposed to obtain this
particular curve and the localization of the optimum power and efficiency along
it. Different tools have been used: \begin{itemize}
	\item \href{http://www.python.org/}{Python} as programming language.
	\item \href{http://numpy.scipy.org/}{Numpy} \cite{Oliphant2006,Dubois1996,Ascher1999}
		/ \href{http://www.scipy.org/}{Scipy} \cite{Jones2001}/
		\href{http://matplotlib.sourceforge.net/}{Matplotlib}
		\cite{Hunter2007} as scientific libraries.
	\item \href{http://openopt.org/Welcome}{OpenOpt} \cite{Kroshko2008} for 
		optimization routines.
\end{itemize} Using equations \eqref{eq:Net_Power} and
\eqref{eq:Thermal_efficiency} allow to evaluate the impact of the different
parameters on the performances of the model, and eventually to optimize it
regarding to power of efficiency. There are two levels of investigation:
\begin{enumerate}
	\item the working temperatures $T_M$ and $T_{m}$.
	\item the physical parameters of the engine  $r_{\Delta s}$, $\epsilon$,
		$\alpha$, $K_{h}$, $K_{l}$, $K_{c}$ and $K_{\text{reg}}$; of the
		gas $\gamma$; and of the surroundings $T_h$ and $T_c$. 
\end{enumerate} A working framework can then be established: \begin{enumerate}
	\item For a fixed set of such parameters, i.e. $r_{\Delta s}$,
		$\epsilon$, and so on: \begin{enumerate}
			\item Computation of the values of $\dot{W}$ and $\eta$
				as a function of $T_{M}$ and $T_{m}$, or conversely
				of $\Delta$ and $\theta$ (see equation
				\eqref{def-delta-theta}), using equations
				\eqref{eq:Net_Power} and \eqref{eq:Thermal_efficiency}.
			\item Optimization of $\dot{W}$ and $\eta$ i.e. find the
				two couples $\left(T_{M},T_{m}\right)$ that
				maximize the power and the efficiency,
				respectively.
			\item Report all the calculated points in power vs. efficiency diagram.
			\item Compute the convex hull in order to obtain the
				shape of this curve, see
				\S\ref{paragraph-Graham}.
		\end{enumerate}
	\item Change the parameters and reassess the steps above, from (a) to
		(d).
\end{enumerate}  This methodology is summarized in
Figure~\ref{fig:Working-framework}. 

\subsection{Convex hull algorithm\label{sub:Convex-hull-algorithm}} \label{paragraph-Graham}

\begin{figure*}
	\centering
	\includegraphics[width=0.75\textwidth]{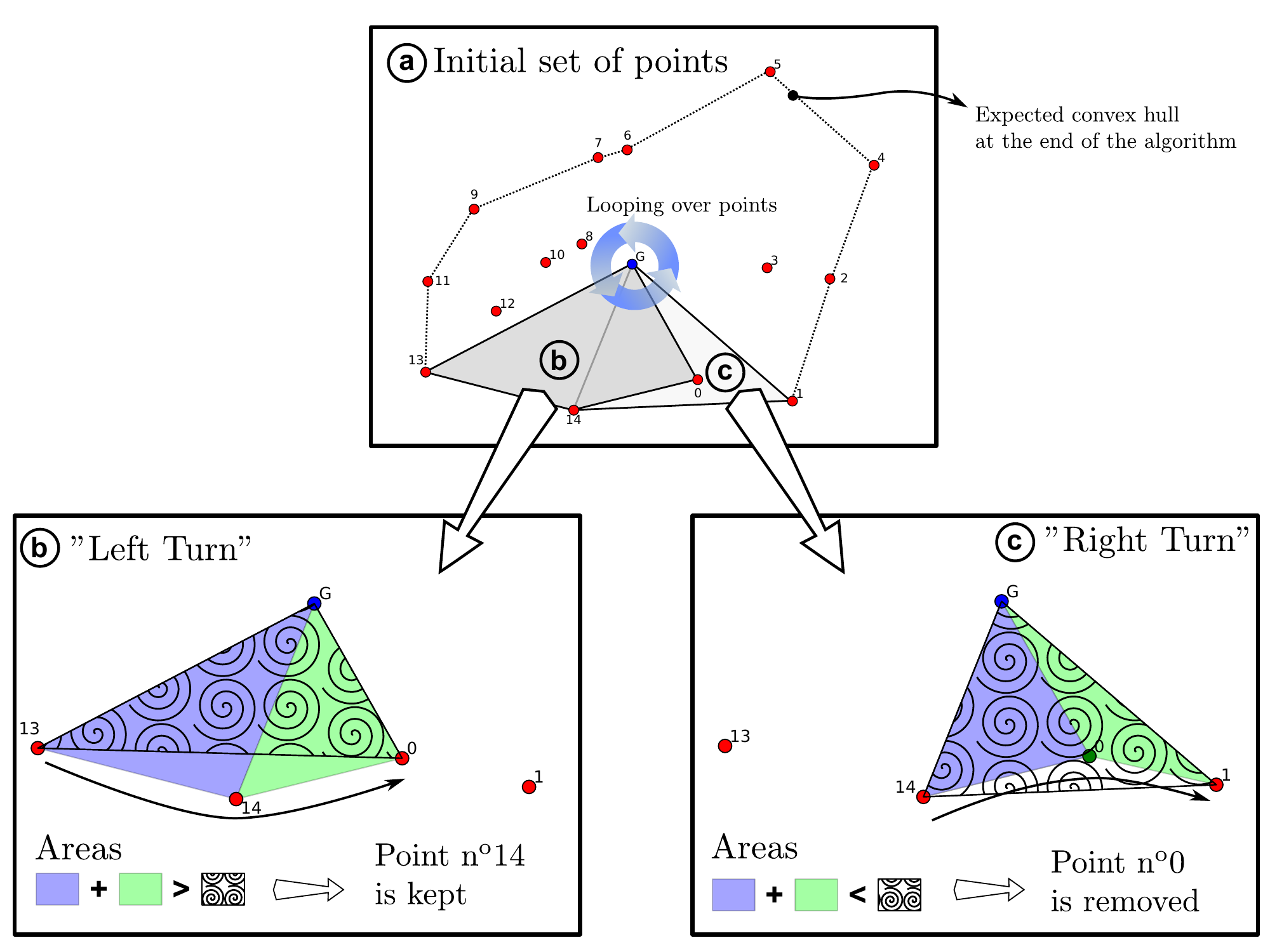}
	\caption{Graham's scan algorithm, used in this paper to compute power
	vs.~efficiency curve of the model, applied to a simpler example of a set
	of $15$ points. Top center: once the position of the center of gravity $G$ is
	known, the counterclockwise angles from $G$ to all these points are
	calculated and sorted. Thus, thanks to comparisons of triangles
	surfaces, a point is either kept as a member of the set enveloppe
	(bottom left) or rejected from the latter (bottom right).}
	\label{fig:Graham-Algorithm}
\end{figure*}

Obtaining the long-awaited power vs. efficiency curve, like the one presented in
Figure~\ref{fig:Power-vs-efficiency}, requires to compute the convex hull (or
convex envelope) of a set of points. The convex hull could be defined as
\emph{the minimal set of points that will define a polygon which includes all
the points of the initial set}. With a large number of points (around $2150$ in
our case) the convex hull is computed thanks to the Graham's algorithm
\cite{Graham1972,Berg2000}. This algorithm can be described as follow:
\begin{enumerate}
	\item Find the center of gravity, noted $G$ of the initial set of
		points.
	\item Compute the counterclockwise angle from $G$ to all other points of
		the set.
	\item Sort the points by angle. 
	\item Construct the boundary by scanning the points in the sorted order
		and performing only “right turns” (trim off “left turns”). 
\end{enumerate} To explain the Graham's scan algorithm, the
Figure~\ref{fig:Graham-Algorithm}-a presents a simpler example with a small
dataset of $15$ points.  The goal is the determination of a ``right turn'' or ``left
turn''. As shown by figure \ref{fig:Graham-Algorithm}-b. A ``left turn'' can be
determined by comparing the areas of several triangles. For a set of 3 points,
\#13, \#14 and \#0 in Figure~\ref{fig:Graham-Algorithm}-b for example, plus the
center of gravity $G$, let's consider 3 triangles ($G-13-14$, $G-14-0$ and
$G-13-0$) and their respective areas. 

If the sum of the surfaces of $G-13-14$ and $G-14-0$ is larger than the surface
of $G-13-0$: the point $14$ is outside, or in terms of directions, the path
$13-14-0$ makes a ``left turn''. The rejection of a point, \#0 in
Figure~\ref{fig:Graham-Algorithm}-c, is decided when a ``right turn'' is
detected. This procedure need to be repeated over all the points, in a
counterclockwise manner, to ensure the convergence.

For a large dataset of about $2150$ points like the one presented in
Figure~\ref{fig:Power-vs-efficiency}, the Graham's scan algorithm could be
easily parallelized by splitting the original dataset. Each processor will deal
with a subset of the original data and a convex hull will be determined for each
subset. By merging the different convex hulls a smaller dataset is obtained on
which the algorithm is applied. The final convex hull will represent the ``power
vs. efficiency'' curve.

\section{Results}

At first, a reference case is considered, with a net power $\dot{W}$ and an
efficiency $\eta$ which are functions of the maximum and minimum working gas
temperatures $T_{M}$ and $T_m$, or conversely of $\Delta$ and $\theta$ and with
a fixed set of parameters $r_{\Delta s}$, $\gamma$, $\epsilon$, $T_{c}$,
$T_{h}$, $\alpha$, $K_{h}$, $K_{l}$, $K_{c}$ and $K_{\text{reg}}$. 

Therefore, optimal temperatures that provided maximum power and efficiency
are determined and secondly an exhaustive investigation of the impact of $5$
parameters, i.e. $\left(\alpha,K_{l},K_{c},K_{h},K_{\text{reg}}\right)$ for a
large range of working gas temperatures is performed.

\subsection{Reference case\label{sub:Reference-case}} \begin{table*}[b!]
	\caption{Fixed and adjustable parameters for equations
	\eqref{eq:Net_Power} and \eqref{eq:Thermal_efficiency}. This particular
	set of parameters is used to obtain
	Figures~\ref{fig:Power-and-efficiency-surface} and
	\ref{fig:Power-vs-efficiency}.}
	\centering
	\begin{tabular}{|c|c|c|c|c|c|}
		\hline 
		& $r_{\Delta s}$ & $\gamma$ & $\epsilon$ & $T_{c}$ & $T_{h}$ \\
		Fixed & - & - & - & $\mathrm{K}$ & $\mathrm{K}$ \\
		& 1.05 & 1.4 & 2 & 300 & 1000\tabularnewline
		\hline 
		& $\alpha$ & $K_{h}$ & $K_{c}$ & $K_{l}$ & $K_{\text{reg}}$ \\
		Adjustable & - & $\mathrm{W \cdot K^{-1}}$ & $\mathrm{W \cdot K^{-1}}$ &
		$\mathrm{W \cdot K^{-1}}$ & $\mathrm{W \cdot K^{-1}}$ \\
		& 0.5005 & 3000 & 3000 & 250 & 10000 \\
		\hline 
	\end{tabular}
	\label{tab:Parameters Ref Case}
\end{table*} 

Although many Stirling engines use different working gases, helium was used in
our study to maintain compatibility with existing standard cycle calculations.
The compression ratio $\epsilon$ and the mass flow was held constant during the
cycle. The gas constant for helium is supposed constant at $r=2080\,\mathrm{J
\cdot kg^{-1} \cdot K^{-1}}$. The entropic parameter $r_{\Delta s}$ is set to
$1.05$. The sink temperature $T_{c}$ is chosen to be at $300\,{\rm K}$ and the
hot source temperature $T_{h}$ at $1000\,{\rm K}$. These values are summarized
in Table~\ref{tab:Parameters Ref Case}.

Hence, for this case the working gas temperatures
\textit{$\left(T_{M},T_{m}\right)$} will be considered to be adjustable within
the range $\left|T_{h}-T_{c}\right|$. For the regenerator effectiveness $\alpha$
and the thermal conductance $K_{i}$, see the lower part of
Table~\ref{tab:Parameters Ref Case}.

\begin{figure*}
	\centering
	\includegraphics[width=\textwidth]{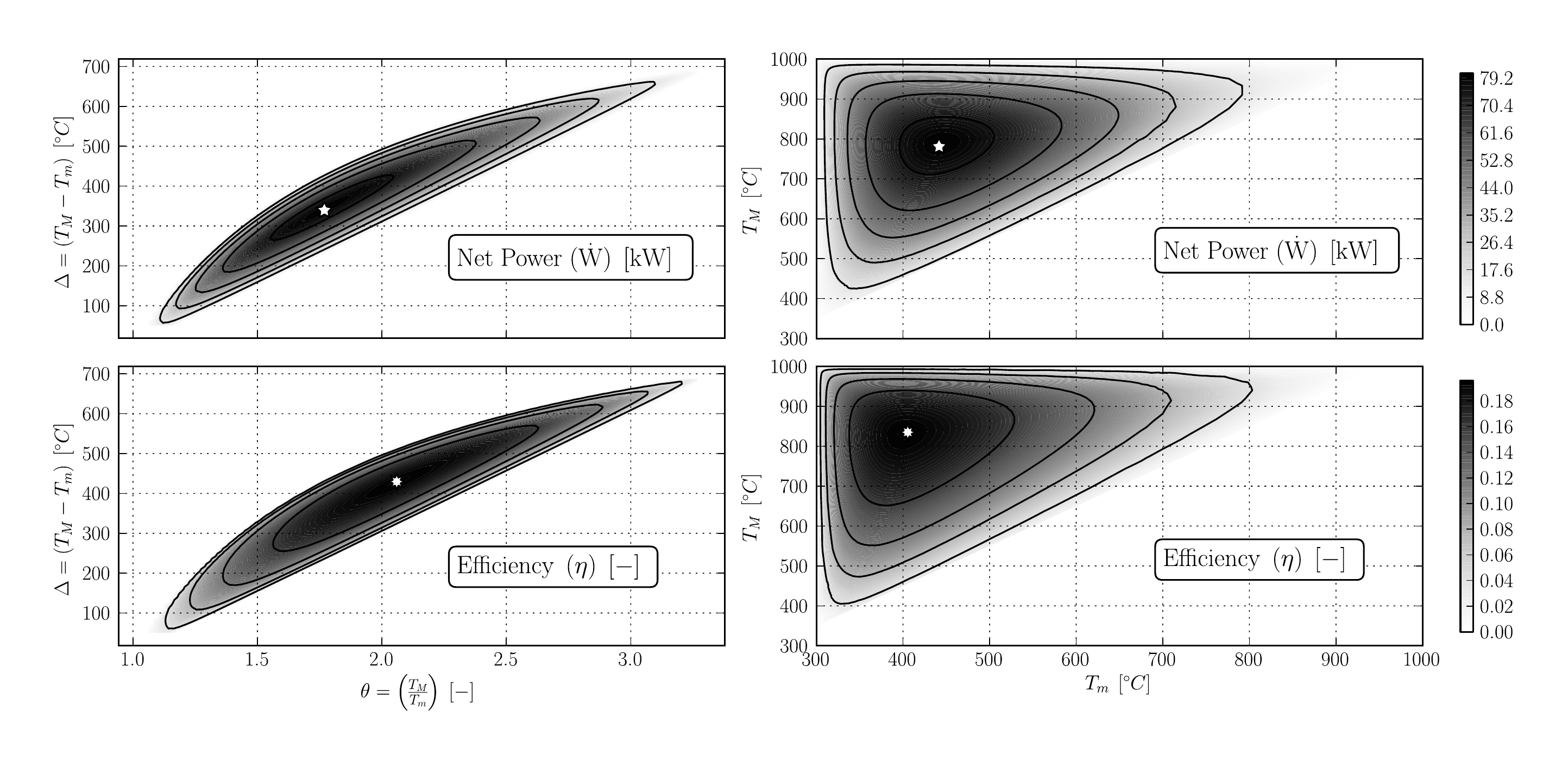}
	\caption{Net power $\dot{W}$ and efficiency $\eta$ as a function
	of $(\theta,\Delta)$ or $(T_M,T_m)$ with the parameters presented in
	Table \ref{tab:Parameters Ref Case}. The star symbols indicate the
	optimum $(\bigstar,*)$.}
	\label{fig:Power-and-efficiency-surface}
\end{figure*}

The surfaces of response obtained thanks to such calculation are presented in
Figure~\ref{fig:Power-and-efficiency-surface}.  The power of equation \eqref{eq:Net_Power} and the thermal
efficiency of equation \eqref{eq:Thermal_efficiency}, respectively presented on
the higher and lower parts of Figure~\ref{fig:Power-and-efficiency-surface}, are
computed regarding to the couple of temperatures $(T_M,T_m)$ on the right part
of the same figure, and to the parameters $(\Delta,\theta)$ on its left part.
For both $\dot{W}$ and $\eta$, an optimum couple of parameters --- $(T_M,T_m)$
and $(\Delta,\theta)$ --- is obtained and drawn, respectively by a symbol
$\bigstar$ and $*$.

\begin{figure*}[t!]
	\begin{center}
		\includegraphics[width=0.75\textwidth]{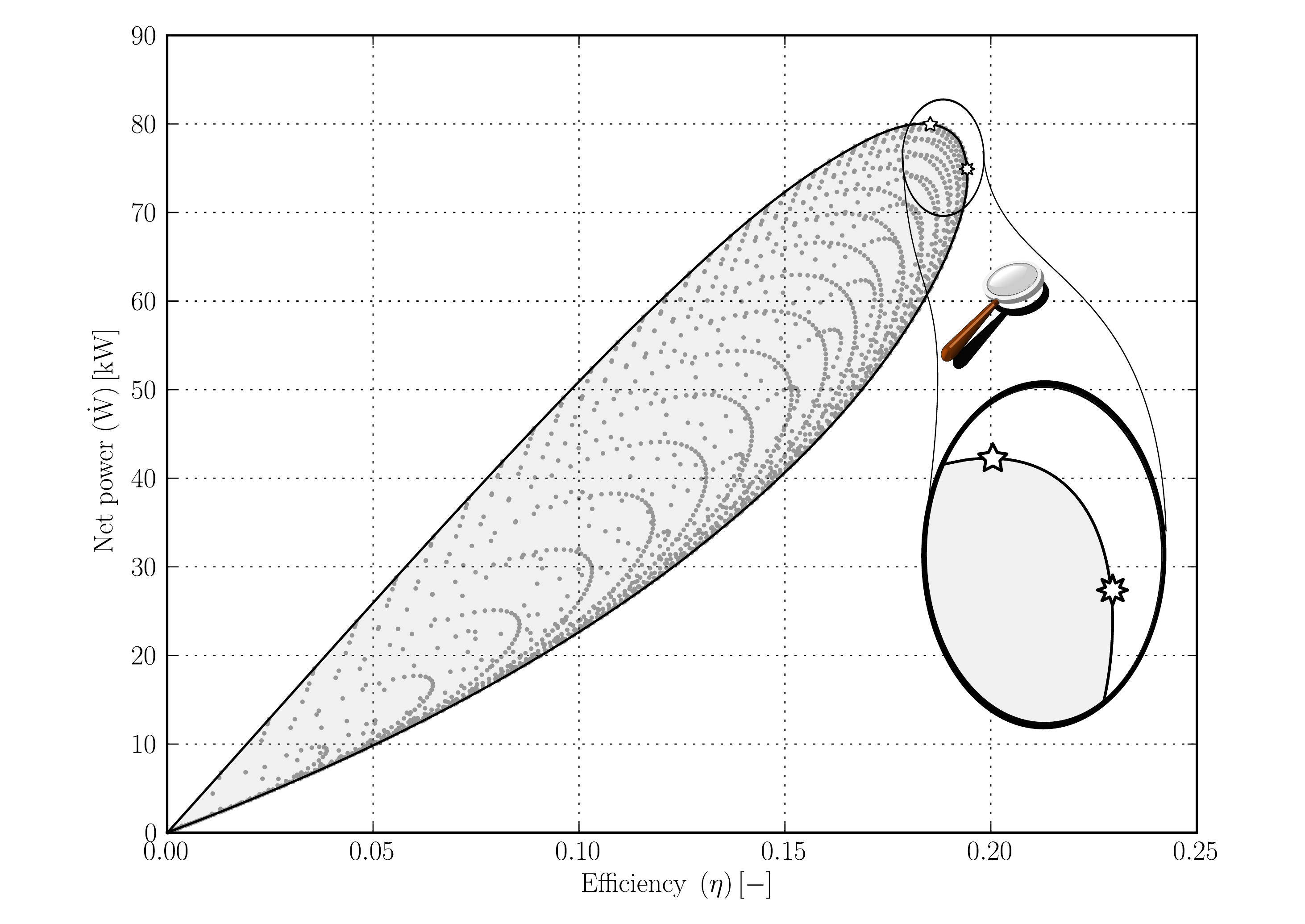}
	\end{center}
	\caption{Net power $\dot{W}$ vs.~efficiency $\eta$ curve with the
	parameters of Table~\ref{tab:Parameters Ref Case}. The symbol $\bigstar$
	indicates the maximum power $\dot{W}_{\max}$ while the symbol $*$
	indicates the maximum efficiency $\eta_{\max}$. Numerical values of
	these optimal parameters are presented in
	Table~\ref{tab:Optimal-values}.} 
	\label{fig:Power-vs-efficiency}
\end{figure*}

Once drawn on the $(\eta,\dot{W})$ plan, the same results gives the ``power vs.
efficiency'' curve of Figure~\ref{fig:Power-vs-efficiency}, on which the maximum
net power $\dot{W}_{\max}$ noted by $\bigstar$ and the maximum efficiency
$\eta_{\max}$ noted by $*$ are represented.  The corresponding temperatures $T_{m}$ and $T_{M}$
are listed in Table~\ref{tab:Optimal-values}. \begin{table*}
	\caption{Optimal values of the working gas temperature for the reference
	case, and the resulting maximum power and efficiency.}
	\centering
	\begin{tabular}{cc|c|c|c}
		&  & Optimum & $T_{m}$ & $T_{M}$ \\
		\hline 
		$\dot{W}_{\max}$ & $\bigstar$ & 80~kW & 442~K & 781~K\\
		\hline 
		$\eta_{\max}$ & $*$ & 0.194 & 405~K & 835~K\\
	\end{tabular}
	\label{tab:Optimal-values}
\end{table*} In terms of temperature, the optimal region is obtained by
maintaining the cold and the hot part of the engine in a narrow range of $\sim
50\,{\rm K}$ with a mean value at $T_m = 423\,{\rm K}$ for the cold side, and
with $T_M = 808\,{\rm K}$ for the hot side.

\subsection{Parametric study\label{sub:Parametric-study}}

A parametric study is conducted in order to observe the effect of the thermal
conductances $K_{h}$, $K_{c}$ and $K_{l}$, and of the regenerator conductance
$K_{\text{reg}}$ and effectiveness $\alpha$ on the values of both maximal power
and maximal efficiency previously obtained on the reference case. For this
study, the fixed parameters remain at their previous values presented in
Table~\ref{tab:Parameters Ref Case} while the adjustable ones may
vary in ranges presented in Table~\ref{tab:Variation-range}. \begin{table*}[b!]
	\caption{Variation ranges considered for the parametric study. Each
	parameter is evaluated 17 times within his own range.}
	\centering
	\begin{tabular}{c|c|c|c|c|c}
		& $\alpha$ & $K_{h}$ & $K_{c}$ & $K_{l}$ & $K_{\text{reg}}$\tabularnewline
		& - & $\mathrm{W.K^{-1}}$ & $\mathrm{W.K^{-1}}$ & $\mathrm{W.K^{-1}}$ & $\mathrm{W.K^{-1}}$\tabularnewline
		\hline 
		minimum & $10^{-3}$ & $10^{3}$ & $10^{3}$ & $10^{-3}$ & $5\cdot10^{3}$\tabularnewline
		maximum & $1$ & $5\cdot10^{3}$ & $5\cdot10^{3}$ & $5\cdot10^{2}$ & $15\cdot10^{3}$\tabularnewline
	\end{tabular}
	\label{tab:Variation-range}
\end{table*} The variation range of each parameter could be seen as wide at
first glance, but it has been chosen to enhance a noticeable change on the net
power and efficiency.

For each set of parameters $\alpha$, $K_{h}$, $K_{l}$, $K_{c}$ and
$K_{\text{reg}}$, the same as procedure as the reference case is followed:
\begin{itemize}
	\item Investigation over $T_{m}$ and $T_{M}$ to obtain a similar result
		as the previous one presented in
		Figure~\ref{fig:Power-and-efficiency-surface} and
		Figure~\ref{fig:Power-vs-efficiency}.
	\item Numerical optimization to determine the maximum power
		$\dot{W}_{\max}$ and the maximum efficiency
		$\eta_{\max}$.
	\item Each optimum result is characterized by a specific couple of
		temperatures $T_{m,\text{opt}}$ and $T_{M,\text{opt}}$.
\end{itemize} A large amount of data has been processed. With 17 levels for each
parameter the total number of cases is $17^{5} \sim 1.4 \cdot 10^{6}$. All the
combinations and the associate calculations were done, but in the present study
only the impact of one parameter at a time is presented, while the others remain
at their reference values presented in Table~\ref{tab:Parameters Ref Case}. 

\begin{figure*}[p]
	\centering
	\includegraphics[width=0.78\textwidth]{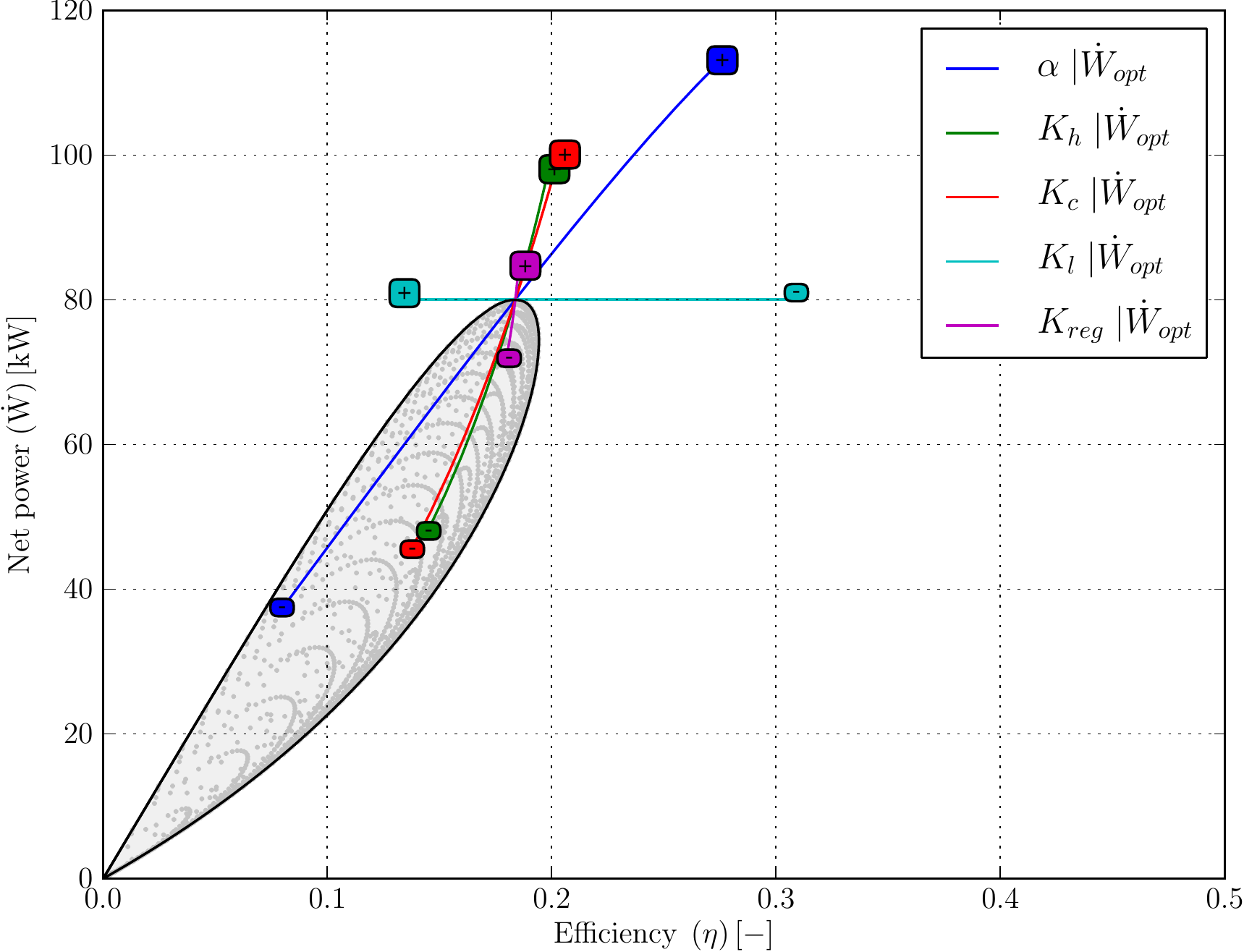}
	\caption{Impact of $\alpha$, $K_l$, $K_c$, $K_h$ and $K_{\text{reg}}$ on
	the optimal net power $\dot{W}_{\text{opt}}$. Each path is marked by a
	\textcircled{+} and a \textcircled{-} to indicate the lower and higher
	value of the selected parameter.}
	\label{fig:Impact-Wopt}
	\bigskip
	\includegraphics[width=0.775\textwidth]{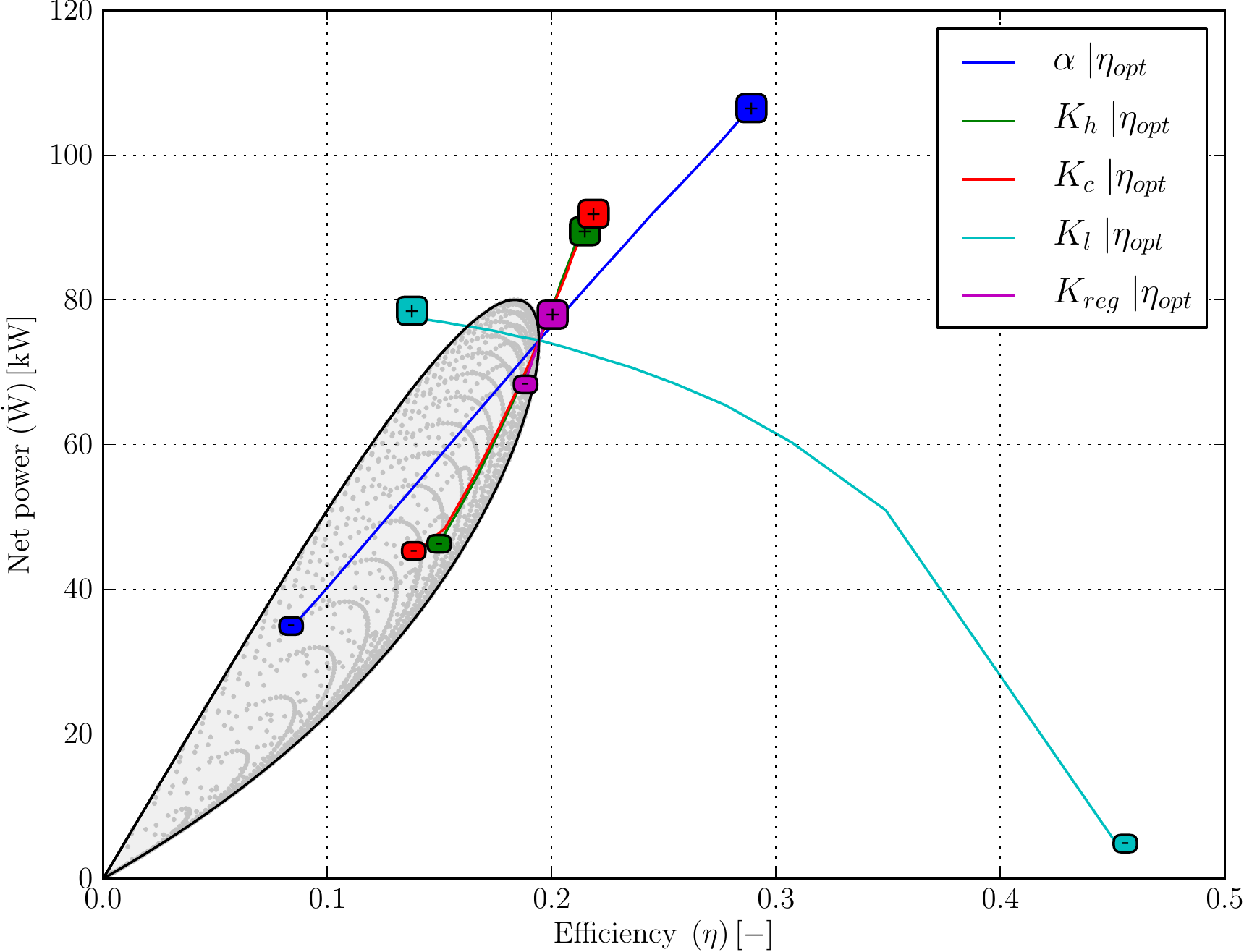}
	\caption{Impact of $\alpha$, $K_l$, $K_c$, $K_h$ and $K_{\text{reg}}$ on
	the optimal efficiency $\eta_{\max}$. Each path is marked by a
	\textcircled{+} and a \textcircled{-} to indicate the lower and higher
	value of the selected parameter.}
	\label{fig:Impact-effopt}
\end{figure*}

In Figures~\ref{fig:Impact-Wopt} and \ref{fig:Impact-effopt}, one could
observe the influence of each parameter on the optimal net power and on the
optimal efficiency, respectively.  

\subsubsection{Influence on optimal net power\label{sub:Optimal-net-power}}

At first, it can be observed in Figure~\ref{fig:Impact-effopt} that an increase
of $\alpha$, $K_{c}$, $K_{h}$ or $K_{\text{reg}}$ enhance both the net power and
the associated efficiency. By increasing those parameters, the heat exchange
that takes place in the engine tends to an ideal behaviour. 

Moreover, from Figure~\ref{fig:Impact-Wopt}, the impact of each parameter
$\alpha \gg K_{c} \sim K_{h} \gg K_{\text{reg}}$ on the optimal net power can be
analyzed. The regenerator effectiveness $\alpha$ seems to be the key parameter
to enhance the performances of the engine. The gain in power is about $45\%$ for
a perfect regenerator, while the loss is greater than $50\%$ for a low value
of $\alpha$. 

The influence of thermal conductances $K_{c}$ and $K_{h}$ on the maximum power
produced are similar, from $-40\%$ to about $+25\%$. These influences are
stronger than the one of the regenerator conductance $K_{\text{reg}}$ which is
almost negligible, from $-10\%$ to $+6\%$.

From a technological and economical points of view, we should keep in mind that
great values of thermal conductances suppose very large heat transfer areas or
very large coefficients of heat exchange. In fact, such configuration would
create much more internal and external irreversibilities and would probably be
unacceptable for a real Stirling engine. 

The increase of the leak conductance $K_{l}$ corresponds to a larger heat
leakage from the hot sink to the cold one. Therefore the efficiency drops
dramatically --- almost divided by 2 --- with a constant optimal net power as
predicted by equation \eqref{eq:Net_Power}. $\dot{W}$ is thus independent of
$K_{l}$. For heat sources of infinite heat reservoirs, the maximum power does
not lead the optimal efficiency, as presented in Figure~\ref{fig:Impact-Wopt}.

\begin{figure*}
	\centering
	\includegraphics[width=0.75\textwidth]{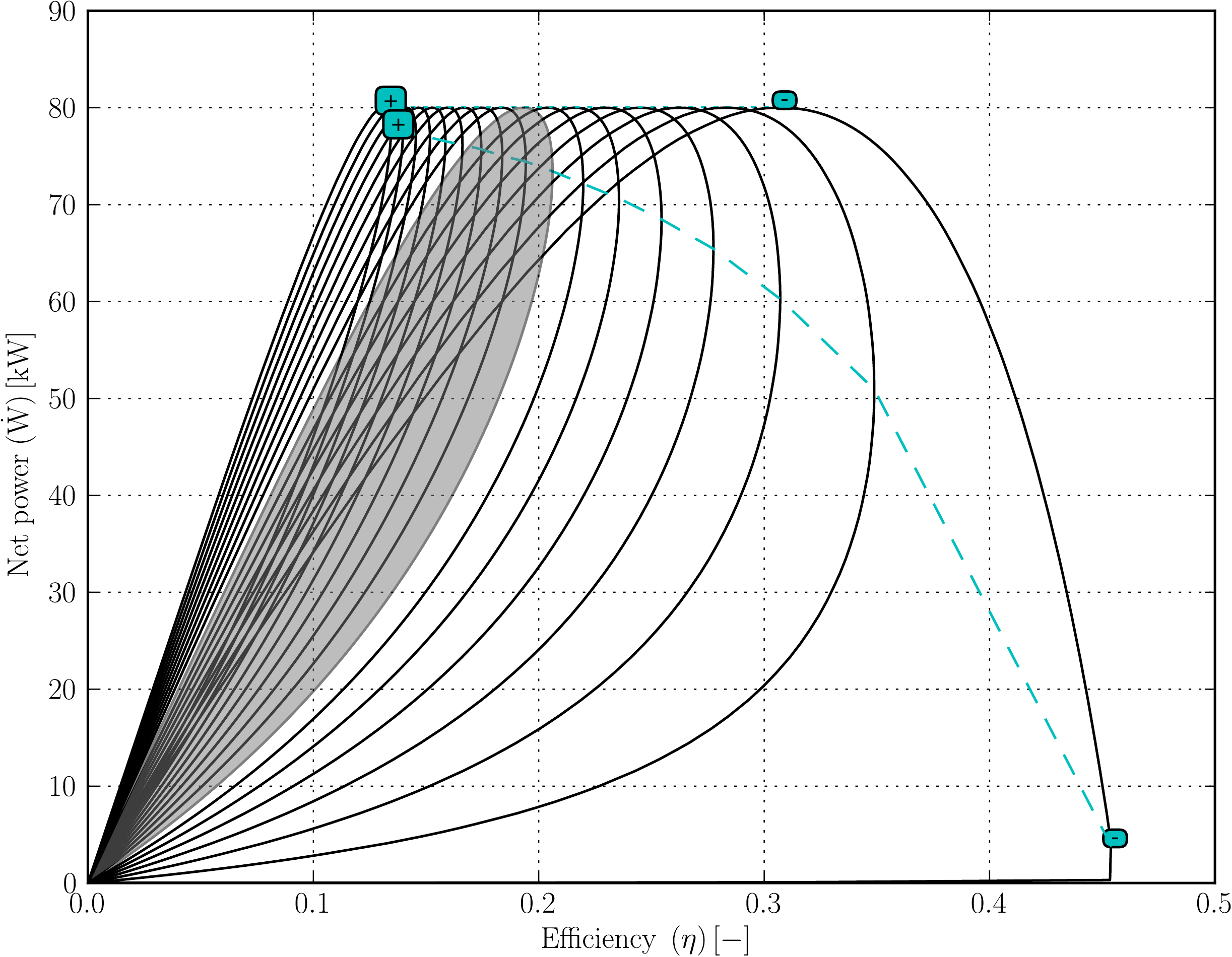}
	\caption{Impact of the leak thermal conductances $K_l$ on the convex
	hull ($-$solid line). Each path is marked by a \textcircled{+} and a
	\textcircled{-} to indicate the lower and higher value. The dashed line
	($--$) corresponds to the change of $K_l$ for the optimal net power,
	same as for Figure~\ref{fig:Impact-Wopt}, and respectively the dotted
	line ($\cdots$) for the optimal efficiency of
	Figure~\ref{fig:Impact-effopt}. The filled grey convex hull is the
	reference case presented in Figures~\ref{fig:Power-vs-efficiency},
	\ref{fig:Impact-Wopt} and \ref{fig:Impact-effopt}.}
	\label{fig:Impact-Kl}
\end{figure*}

\subsubsection{Influence on optimal efficiency\label{sub:Optimal-efficiency}}

In Figure~\ref{fig:Impact-effopt}, it can be observed that the parameters
$\alpha$, $K_{c}$, $K_{h}$ and $K_{\text{reg}}$ have the same influence on the
maximum efficiency $\eta_{\max}$ that they have on the maximum power
$\dot{W}_{\max}$, as presented in Figure~\ref{fig:Impact-Wopt}. 

About the heat leakage conductance $K_{l}$, the conclusion is not so obvious
because an increase of $K_{l}$ both leads to a dramatic loss of the maximum
efficiency value and to a relative increase of the net power.  The
Figure~\ref{fig:Impact-Kl} is used for a better understanding of this
phenomenon. The ``power vs. efficiency'' curves are heavily altered by the
$K_{l}$ values, going from a lens-shaped curve for high values of $K_{l}$ to a
parabolic shaped curve for small values of $K_{l}$. In fact, when $K_{l}$ tends
to zero the system tends to the endoreversible engine, as stated in
\S\ref{sub:Comp-with-endoreversible}. The opposite case with an increase of
$K_{l}$ tends to an Stirling engine which has the ability to produce power but
with a very low efficiency. This phenomenon can be explained by the ideal sinks
used, of infinite capacities at temperatures $T_{c}$ and $T_{h}$. For any value
of $K_{l}$ --- even a large one --- some heat rate will flow through the
Stirling engine. These rates produce a noticeable net power but are negligible
in comparaison with the main heat leakage that take place between the hot and
cold sink.

\section{Conclusion}

A thermodynamic analysis of an irreversible realistic Stirling engine model has
been carried out and maximum power and maximum efficiency have been obtained
numerically in order to provide an estimate of the potential performance bounds
for real Stirling engine. 

A fully detailed theoretical model is obtained and studied through an innovative
method in the Finite Dimension Thermodynamic field. The use of massive
computations allows to predict the behaviour of the engine for a large range of
its own specific properties. The power and efficiency are easily obtained by
using a computational geometry algorithm, the called Graham's algorithm, that
determines the convex hull around the obtained values of the former.  This study
has shown that the final envelop is fully determined by a set of parameters
which are the overall conductances on the hot and cold side and the regenerator
properties, i.e. its conductance and effectiveness.

Parametric studies have been conducted with respect to the effect of reservoir
temperature, regenerator effectiveness, entropic parameter and reservoir heat
conductances. If maximum power and maximum efficiencies increase with the heat
source temperature, the different sources of irreversibilities result in a
decrease on the performances of the Stirling cycle engine. We demonstrated the
key role of the regenerator effectiveness on the performances of the Stirling
engine in terms of net power and efficiency. The results of this paper can be
used to provide additional insight into the design of irreversible Stirling
machine with imperfect regeneration like engines, refrigerators, heat pumps and
could be extended to Ericsson machines and any machines with regenerative
processes. Additionally the methodology used in this study could be applied in
any Finite Dimension Thermodynamic analysis where the analytical expressions are
too mathematically complex to be manipulated. Further work should be done to
enhance the methodology to be usable to compare the entropy production to the
relationship between the maximum power and efficiency and to characterize a
dynamic model of a Stirling engine.
\section*{Nomenclature}

\begin{tabbing}
	\hspace{1cm} \= \hspace{7cm} \kill
	\> {\bf Notations} \\
	$A$ \> Area, $[{\rm m}^2]$ \\
	$c_v$ \> Specific heat at constant volume, $[{\rm J \cdot kg}^{-1} \cdot
	{\rm K}^{-1}]$ \\
	$F$ \> Form factor \\
	$h$ \> Heat transfer coefficient, $[{\rm W \cdot m}^{-2} \cdot {\rm
	K}^{-1}]$ \\
	$K$ \> Thermal conductance, $[{\rm W/K}]$ \\
	$m$ \> Mass, $[{\rm kg}]$ \\
	$p$ \> Pressure, $[{\rm bar}]$ \\
	$\dot{Q}$ \> Heat rate, $[{\rm W}]$ \\
	$r$ \> Specific constant of the gas, $[{\rm J \cdot kg}^{-1} \cdot {\rm
	K}^{-1}]$ \\
	$r_{\Delta s}$ \> Internal irreversibility coefficient \\
	$S$ \> Entropy, $[{\rm W \cdot K}^{-1}]$ \\
	$t$ \> Time, $[{\rm s}]$ \\
	$T$ \> Temperature, $[{\rm K}]$ \\
	$V$ \> Volume, $[{\rm m}^3]$ \\
	$\dot{W}$ \> Mechanical power, $[{\rm W}]$ \\
	\> {\bf Greek symbols} \\
	$\alpha$ \> Regenerator effectiveness \\
	$\Delta$ \> Difference $T_M - T_m$ \\
	$\epsilon$ \> Compression ratio of the engine \\
	$\eta$ \> Energy efficiency \\
	$\theta$ \> Ratio $T_M/T_m$. \\
	$\sigma$ \> Stefan-Boltzmann constant \\
	$\gamma$ \> Adiabatic index \\
	\> {\bf Subscripts} \\
	$c$ \> Cold sink \\
	endo \> Endoreversible engine \\
	$h$ \> Hot source \\
	$l$ \> Heat leak \\
	$m$ \> Minimum temperature of working gas \\
	$M$ \> Maximum temperature of working gas \\
	rev \> Reversible engine \\
	\> {\bf Acronyms} \\
	FDT \> Finite Dimension Thermodynamics \\
	FTT \> Finite Time Thermodynamics \\
\end{tabbing}

\bibliographystyle{unsrtnat}
\bibliography{biblio-arxiv}

\end{multicols}
\end{document}